\documentclass[prb, showpacs, superscriptaddress]{revtex4}

\usepackage{graphicx}%

\usepackage{amssymb}
\usepackage{amsmath}
\usepackage{mathtools}

\begin{document}

\title{Close packed structure with finite range interaction: computational mechanics of layer pair interaction}

\author{Edwin \surname{Rodriguez-Horta}}
\author{Ernesto \surname{Estevez-Rams}}
\affiliation{Facultad de F\'isica-IMRE, Universidad de la Habana, San Lazaro y L. CP 10400. C. Habana. Cuba}
\email{estevez@fisica.uh.cu}

\author{Reinhard \surname{Neder}}
\affiliation{Kristallographie und Strukturphysik. Universit\"at Erlangen. Germany}

\author{Raimundo \surname{Lora-Serrano}}
\affiliation{Universidade Federal de Uberlandia, AV. Joao Naves de Avila, 2121- Campus Santa Monica, CEP 38408-144, Minas Gerais, Brazil}

\date{\today}

\begin{abstract}
The stacking problem is approached by computational mechanics, using an Ising next nearest neighbor model. Computational mechanics allows to treat the stacking arrangement as an information processing system in the light of a symbol generating process. A general method for solving the stochastic matrix  of the random Gibbs field is presented, and then applied to the problem at hand. The corresponding phase diagram is then discussed in terms of the underlying $\epsilon$-machine, or optimal finite state machine, describing statistically the system. The occurrence of higher order polytypes at the borders of the phase diagram is also analyzed. Discussion of the applicability of the model to real system such as ZnS and Cobalt is done. The method derived is directly generalizable to any one dimensional model with finite range interaction.
\end{abstract}


\maketitle

\section{Introduction}

Crystals are the epitome of order in nature. In the ideal case, the knowledge of the atomic arrangement in a finite-size volume allows to unambiguously calculate  the atomic arrangement at any other position in the solid. That is, at the end, the very idea of long range order. Yet, real crystals are far from such idealization.  Randomness disrupt order to some extent in every real crystal. When the density of randomness is low, it can be treated as a perturbation of the underlying perfect order. Such cases have been by large the most dealt with, and there is an extensive body of literature covering models of perturbation disorder in crystals. Such models proceed from classifying different types of order disruption as 'defects' of the otherwise ideal crystal. Vacancies, dislocations and planar faulting are then recognized as defects, and their occurrence is characterized in different ways. Under this point of view, models of disorder are largely models of different types of defects or a combination of them. It is then difficult to characterize disorder in a broader sense, regardless of the actual type of defect or even the case where it is hard to define what the underlying perfectly ordered structure is.

This last case is not rare, it happens in a vast number of crystals when they undergo reconstructive phase transitions, changing from some crystal structure to another \cite{toledano96}. Such is the case, for example, for close packed structures as they change from one polytype order to another. Layer crystals that differ in the order of arrangement of the otherwise identical layers are called a polytype family. For example, face centered cubic  and hexagonal close packed cobalt belong to the same polytype family. Polytipism is ubiquitous in a large number of close packed structures, being silicon carbide and zinc sulphide compounds archetypal examples. Polytipism is also present in more simple systems such as pure Cobalt or Lithium. Closely related to polytypism is the occurrence of extensive stacking disorder. When the energy difference between two stacking arrangements is small, disorder can be the result of thermal noise or entropic-driven lowering of the system free energy. If phase transition is considered a continuous process of layer ordering rearrangement, then at some point, it will be hard to assert, even in terms of the starting and ending structures, if a particular type of planar defect is happening or even if it can be defined. If we still try to explain the reordering process in terms of planar defects,a particular narrow view could be forced to a process amenable to a broader and more comprehensive approach. 

Borrowing tools from other particular fields of physics and mathematics could be a way of approaching disorder in crystals in a broader way. After all, pattern, order and randomness are concepts at the root of several scientific areas, that include statistical mechanics, thermodynamics, non-lineal dynamics and information theory, among others \cite{crutchfield12}. In such fields, instead of seeking particular models of disorder, unpredictability is first characterized by using global measures, such as entropy in its different flavors, and entropy derived magnitudes. This is not new for crystallography, however, this discipline has been slow in using such approach in a  more systematic way as has been done in those other fields. This new ideas have greatly  enhanced our set of tools to conceptually and experimentally understand randomness, pattern and order. Computational mechanics, a field of complexity derived directly from Shannon information theory, is one of these emerging areas \cite{feldman03}.

Computational mechanics, pioneered by the group of Crutchfield \cite{crutchfield89}, attempts to look at the emergence of patterns and there relation with disorder by considering a physical system as a natural probabilistic 'computational machine' with a given capacity to input, store, process and output information. This machines can be classified by the hierarchy of language types they can process, as has been done in computer science. Once the least sophisticated machine capable of optimally reproduce the physical system is found, its informational capacity of storage, processing and its amount of irreducible disorder can be quantified in a precise manner. Different systems can then be compared by the hierarchy of the machines used, their topology, and the resources needed (for a more recent review the reader can refer to \cite{crutchfield12}).

From the general framework of computational mechanics Varn and coworkers have been recasting the analysis of planar disorder in layer structures \cite{varn02,varn04,varn06,varn13,varn13a,riechers14}, within a more ambitious goal of looking into disorder in solids from this new perspective.  They have successfully used such framework to relate diffraction pattern with the optimal finite state machine capable of statistically describing the system. 

As far as the authors know, no attempt has been done to study polytypism from the computational mechanics point of view. This paper will report the results of such analysis for a common type of Ising interaction model used before in polytype studies. In a second paper this analysis will be carried out with a different interaction model. In both cases a particular Hamiltonian with local interaction will be considered, and the implications when the control parameters of the model vary in a meaningful range of values will be studied. The model presented here is not new, on the contrary, it is the most widely studied model of interaction for close packed structures \cite{uppal80,kabra88,shaw90}. What is new is the use of computational mechanics, which will allow to characterize the emergence of pattern and disorder in a more sound way. Computational mechanics also allows to discuss the emergence of long order polytypes and their probability of occurrence, as well as to determine when a given polytype can be intrinsic to the thermodynamics of the system and when it is the result of arrested states at the boundaries between ordered blocks.

The paper is organized as follows. We start in Section \ref{sec:cps} by giving some background on the subject of close packed structures and clarifying the notation used. In Section \ref{sec:markov} the mathematics of the Markov process describing the layer ordering, the used Ising model, and the computational mechanics approach developed through the $\epsilon$-machine construction is described. In Section \ref{sec:NNN} the general developed framework will be applied to the next nearest neighbor one-dimensional (1D) Ising model. The phase diagram will be discussed in terms of the underlying finite state machine describing the dynamics of the system. Discussion of the results will follow in Section \ref{sec:discussion} and from there conclusions will be drawn. The more involved mathematical deductions are left for the appendix.

\section{Definition and notation.\label{sec:cps}}

Close packed structures (CPS) are built by stacking hexagonal layers in the direction perpendicular to the diperiodic layer. They are a particular type of OD (Order-Disorder) structures \cite{durovic97}. In this case, the layers can only be found in three positions perpendicular to the stacking direction \cite{durovic97,pandey2}. If each position is labeled with a different letter $A$, $B$ and $C$, the close packed condition means that two layers  which bear the same letter can not occur consecutively. In physical terms two consecutive layers that exactly overlap in the projection along the stacking direction are forbidden.  In terms of the three letters, the ideal face centered cubic ($FCC$), hexagonal compact ($HCP$) and double hexagonal compact ($DHCP$) are described by $ABC\ldots$, $AB\ldots$ and $ABCB\ldots$, respectively \cite{verma66}. 

The description of stacking ordering using the $A$, $B$ and $C$ characters is redundant due to the close packed constraint \cite{verma66,triguniyat91}. The actual labeling of a layer by a given letter representing a layer displacement is arbitrary. Consider each letter as representing a displacement described by a certain number of times a constant vector $\vec{s}$, perpendicular to the stacking direction. Let $A$ represents $0 \vec{s}$, $B$ represents $1 \vec{s}$ and $C$ the displacement $2 \vec{s}$. Displacing each layer in the ordering by the same integer number of times the $\vec{s}$ vector can change the letters from one to another (e.g $A\rightarrow B \rightarrow C$) without changing the physical nature of the crystal. It just represents a shift in the origin or, equivalently, a translation of the crystal as a whole. It is then clear that what makes sense from the physical point of view, is the relative displacements of one layer with the other layers in the crystal. Such relative displacement for consecutive layers can be coded using a binary alphabet: a symbol for two consecutive layers with relative displacement $\vec{s}$ and another symbol for a relative displacement of $2 \vec{s}=-\vec{s}$. This coding is known under several names \cite{verma66}, being the  H\"{a}gg code and the Nabarro-Frank code two of the most used names and symbols. The first  uses $+$ and $-$ as symbols for the two described displacements and the latter uses instead symbols $\bigtriangleup$ and $\bigtriangledown$. A binary number representation can also be given and will be used here with $-1$ for $\bigtriangledown$ and $1$ for $\bigtriangleup$. There is a one-to-one relation between the binary code and the $ABC$ coding up to the labeling of the first layer. The stacking arrangement is now represented by a binary string $s_{1} s_{2} s_{3} \ldots s_{i} \ldots s_{N}$. Each $s_{i}$, called the spin at position $i$, represents the displacement of two consecutive layers at positions $i$ and $i+1$.

It is of common use, to shorten the binary representation of a polytype by using a run length encoding algorithm that is known in crystallography as the Zhdanov symbol. In the Zhdanov representation, consecutive runs of the same symbol, in the binary stacking description, is represented by a number giving the length of the run \cite{verma66}. For example, the polytype $\bigtriangleup\bigtriangleup\bigtriangleup\bigtriangledown\bigtriangledown\bigtriangledown \bigtriangledown \bigtriangledown\bigtriangledown$ has Zhdanov symbol $36$. If a fragment of the spin ordering repeats itself, then in the Zhdanov symbol it appears between parentheses with a subscript showing the number of repetitions; e.g. $\bigtriangleup\bigtriangledown\bigtriangledown\bigtriangleup\bigtriangledown\bigtriangledown\bigtriangleup\bigtriangledown\bigtriangledown$ corresponds to the Zhdanov symbol $(12)_3$.

Once the stacking sequence in a CPS is coded as a binary sequence, several phenomena occurring in such crystals can be approached as a mathematical problem over the binary code. That has been the case for the description of polytypism in CPS structures \cite{mclarnan,iglesias81,estevez05,estevez05a,iglesias06a,iglesias06b}. 

It is more or less straightforward to study polytypism and stacking disorder by writing a Hamiltonian that describes the interaction between the binary codes, treated as spins, and then to analyze the system as a case of spin interaction. This interaction can be long range, finite range or both \cite{uppal80,kabra88,shaw90}. 

\section{Representing a layer ordering as a Markov process.\label{sec:markov}}

Our goal is to represent the stacking ordering of $N$ layers as the output of a Markov process. A Markov process is defined by the alphabet of the output symbols; a set of states $\mathcal{S}=\{\eta_1, \eta_2, \eta_3, \ldots \eta_t\}$; the initial probability distribution $\mu_0=\{Pr(\eta_1), Pr(\eta_2), \ldots Pr(\eta_t)\}$, where each entry is the probability that the Markov process will start at a given state; and the set of transition probability between states when a particular symbol in the alphabet is output. When the arriving state unambiguously determines the output symbol, then a single matrix  will determine all transitions, this matrix $P$ will be called the stochastic matrix and has the form
\begin{equation}
P=\left ( \begin{array}{cccc}
 Pr(\eta_1|\eta_1) &  Pr(\eta_2|\eta_1) & \ldots & Pr(\eta_t|\eta_1) \\\\
 Pr(\eta_1|\eta_2) &  Pr(\eta_2|\eta_2) & \ldots & Pr(\eta_t|\eta_2) \\\\
 \ldots & \ldots & \ldots & \ldots\\\\
 Pr(\eta_1|\eta_t) &  Pr(\eta_2|\eta_t) & \ldots & Pr(\eta_t|\eta_t) \\\\
 \end{array}\right ),\label{eq:hmm}
\end{equation}
where each entry $Pr(\eta_i|\eta_j)$ is the probability of making a transition from state $\eta_j$ to state $\eta_i$.

By construction, the Markov process can de described by a finite state automaton (FSA) graphically represented by a directed graph, where each node is a state $\eta_i$ and each vertex connecting two nodes is directed in the sense of a transition from a state $\eta_i$ to a state $\eta_j$ and labeled by $Pr(\eta_j| \eta_i)$. The less sophisticated Markov process capable of statistically reproducing the stacking ordering is a case of what in computational mechanics is called an $\epsilon$-machine. Less sophisticated means that it uses the least amount of memory related to the total number of states.  

$\epsilon$-machines are at the heart of computational mechanics. They represent, as optimally as possible, computational machines that statistically generate the stacking ordering. It is in this sense that characterizing the $\epsilon$-machine amounts to characterizing the dynamics that causes the layer to order in a certain type of stacking sequence.

It is useful at this point to introduce bras $\langle \bullet | $, and kets $| \bullet \rangle$ to denote row and column vectors, respectively. It is clear then, that $\langle \bullet | \bullet \rangle$ will be a scalar resulting from the multiplication of a row with a column vector (scalar product). 

In what follows, the kind of layer orderings that will be studied will be those resulting from local type interactions that can be represented as an Ising model. The first step is then to translate the Ising model into the described Markov process. But first, the Ising models of interest will be formally introduced.

\subsection{Ising model for layer interaction in CPS.\label{sec:ising}}

When considering only finite range interactions, a large class of systems can be cased in the framework of the Ising model, which has a Hamiltonian of the type:
\begin{equation}
 \displaystyle \mathcal{H}=-B \sum\limits_{i=1}^{N} s_{i}-\sum\limits_{k=1}^{n}J_{k}\sum\limits_{j=1}^{N-n}s_{j}s_{j+k},\label{eq:isinghamiltoniana}
\end{equation}
where $B$ is an external field. The $J_k$ is the interaction parameter for pair of spins separated by $k$ number of layers. The interaction is taken to be at most between spins separated by $n$ layers, the maximum interaction range, which is the upper limit in the sum over $k$. $s_i$ represents the spin at position $i$ in the layer ordering. The total number of layers is $N$. It must remembered that spins do not represent layers directly, but instead, pair of layers. 

In the case of stacking arrangement $B$ is usually made zero resulting in  
\begin{equation}
 \displaystyle \mathcal{H}=-\sum\limits_{k=1}^{n}J_{k}\sum\limits_{j=1}^{N-n}s_{j}s_{j+k}.\label{eq:isinghamiltonian}
\end{equation}
If $S^{N}=s_1 s_2 s_3, \ldots, s_N$ is a given sequence of spins describing a stacking ordering then, $\mathcal{H}$ is a function of $S^N$ ($\mathcal{H}=\mathcal{H}(S^N)$).

As far as the authors know, M. K. Uppal \cite{uppal80} made the first attempt to study polytypism within the framework of the Ising model. In their study they performed Monte Carlo simulations combining both short and infinite range interactions, all of them of pairwise type. V. K. Kabra and D. Pandey further developed the Ising approach to polytypism using up to a range three interactions \cite{kabra88}. It was in the context of their work that correlation length was introduced as a measure of the number of layers over which a disordered stacking arrangement loses memory. J. J. A. Shaw and V. Heine \cite{shaw90} thoroughly studied polytypism in SiC using an Ising model more general than equation (\ref{eq:isinghamiltonian}), which incorporates terms involving the product of an even number of spins. The first of such terms appears for an interaction range of $4$. In all these cases, the  H\"{a}gg coding is used to cast the stacking arrangement into a binary code or string. 

When a Hamiltonian is introduced acting over a system coded as a finite alphabet string, the whole system can be seen as a dynamical system where, as a result of the different interaction terms, patterns and disorder can arise in the sequence of characters. This is a common approach in complexity analysis. Possible goals in such study could be the identification of pre-known patterns within the string, or the discovery of patterns with no a-priory knowledge of which regularity should be expected. Also, the ordering of patterns and the assessment of the amount of irreducible disorder in the string are possible questions to be answered. 

Ising models, as the one described by the Hamiltonian (\ref{eq:isinghamiltonian}), are long known in statistical physics. The analytical solution for such model follows from the calculation of the partition function $Z_{N}$:
\begin{equation}
 Z_{N}=\sum\limits_{\{S^N\}} \exp[-\beta \mathcal{H}(S^N)],\label{eq:part}
\end{equation}
where the sum is carried over all possible spin configurations of length $N$ (there will be at most, $2^N$ of such configurations). $\beta$ is the Boltzmann factor ($=1/k_B T$).

Knowing the partition function allows to calculate the equilibrium thermodynamic quantities which globally characterize the system \cite{lavis99}. This procedure has been well understood for many years. It is an important part at the core of statistical mechanics (See for example \cite{greiner95}).

A partition can be made of the spin configuration in non-overlapping blocks of length $n$, the maximum interaction range in the Hamiltonian (\ref{eq:isinghamiltonian}), and the $m^{th}$-block of spins is denoted by $\eta_{m}$. If $s^{(m)}_i$ represents the $m\times n+i$ spin in the stacking arrangement then the $\eta_m$ block is formed by the $n$ spins
\begin{equation*}
 \eta_{m}=s_1^{(m)}s_2^{(m)}\ldots s_{n}^{(m)}.
\end{equation*}
There are $2^{n}$ different blocks $\eta$. The spin configuration can be written as a sequence of $L(=N/n)$ $\eta$ blocks.

Define a $|U\rangle$ vector of $2^{n}$  components by
\begin{equation}
\displaystyle u_{\eta_p}=\exp\left (-\frac{1}{2}\beta \sum\limits_{i=1}^{n-1}\sum\limits_{k=i+1}^{n}J_{k-i}s^{(p)}_i s^{(p)}_k \right ),\label{eq:uv}
\end{equation}
and a $2^n \times 2^n$  matrix $V$ with entries
\begin{equation}
 v_{\eta_p \eta_{p+1}}=u_{\eta_p}u_{\eta_{p+1}}\exp \left ( -\beta \sum\limits_{i=1}^{n} \sum\limits_{k=1}^{i} J_{n+k-i} s_{i}^{(p)}s_{k}^{(p+1)}\right ).\label{eq:vm}
\end{equation}
In terms of these two operators the partition function (\ref{eq:part}), using equation (\ref{eq:isinghamiltonian}) can be rewritten as
\begin{equation}
 Z_{N}=\langle U | V^{L-1} | U \rangle.\label{eq:partition}
\end{equation}
In statistical mechanics $V$ is known as the transfer matrix \cite{dobson69}.

Feldman \cite{feldman98,crutchfield97} proved that the  Ising model can be casted into a Markov process where the states are the $\eta$ blocks and the stochastic matrix can be derived from the transfer matrix. In appendix A we show a general derivation of the corresponding expressions.

The fact that the Ising model can be described as a Markov process with a given stochastic matrix with entries $Pr(\eta_{i}|\eta_{j})$, defines a computational machine that generates and stores information. The output symbols of these machines, as transition is made from any state to state $\eta_i$, will precisely be the spins that define the arriving $\eta_i$ block. By construction, the computational machine behind the Ising model is sequential: it generates the layer arrangement in sequential order. This last fact is by no means trivial. The Ising model is an example of a Gibbs random field, where the probability of a spin at position $i$ taking a particular value, depends on a local neighborhood, both to the right and to the left of the $i$ layer. Therefore, spin values at all positions are simultaneously determined by local configurations whose union spans the whole stacking ordering.

\subsection{The $\epsilon$-machine.\label{sec:em}}

It is defined that all $\eta_{j}$ blocks that condition the same probability distribution (over all future $\eta_{i}$) belong to the same causal state ($C_{p}$). A causal state is therefore a set of $\eta$ blocks that probabilistically determines the same future \cite{shalizi01}. The knowledge about in which causal state the computational machine is determines, as optimally as possible, the future outcomes of the process:  the set of causal states uniquely determines the future of a sequence. The set of all causal states will be denoted by $\mathcal{C}$ with cardinality $|\mathcal{C}|$. Two blocks belonging to the same causal state $C_{p}$ define identical types of rows in the stochastic matrix. The $\epsilon$-machine is then the one derived from the Markov process with nodes given by the causal states and the corresponding stochastic transition matrix.

The idea that causal states captures is that of causality in a broad sense. In the case of layer arrangement, this causality originates in the fact that the last $n$ spins, that is, the last $\eta$ block, is as much memory as needed to predict optimally the next spins outcomes. Knowing a longer string of spins does not improve the predictive capacity. However, causal states adds an additional key aspect of causality: $\eta$ blocks belonging to the same causal state determine the same spin outcomes in a statistical sense. Therefore, in terms of predictability, it can be redundant to account for each $\eta$  block; the knowledge of the causal states suffice to achieve the optimal predictability.

The probability of a causal state will be by definition
\begin{equation}
 \displaystyle Pr(C_{p})=\sum\limits_{\eta_{j}\in C_{p}} Pr(\eta_{j}).\label{eq:pr}
\end{equation}
The statistical complexity is defined as the Shannon entropy \cite{arndt01} over the causal states
\begin{equation}
 \displaystyle C_{\mu}= - \sum\limits_{C_{p}\in\mathcal{C}}Pr(C_{p})\log Pr(C_{p}),\label{eq:Cmu}
\end{equation}
logarithm base two is taken base and therefore, units are bits. Being a function of the causal states probability, $C_{\mu}$ is the entropy related to the memory of the system \cite{feldman98}.

The entropy density is defined to assert the amount of randomness that can not be accommodated for in any way. It is the limit in the infinite string of the Shannon entropy over the spin sequences divided by the sequence length. In the case of the Ising model it can be calculated from \cite{feldman98}:
\begin{equation}
 \displaystyle  h_{\mu}=-\sum\limits_{C_{\alpha}\in \mathcal{C}} Pr(C_{\alpha})\sum\limits_{\{\eta_{k}\}}Pr(\eta_{k}|C_{\alpha})\log Pr(\eta_{k}|C_{\alpha}),\label{eq:entropyrateblock}
\end{equation}
where $P(\eta_k| C_\alpha)$ is the conditional probability of emitting a $\eta_k$ block as a transition is made from the $C_\alpha$ causal state. As already explained $h_{\mu}$ is the irreducible randomness of the stacking sequence, a measure of the amount of the layer ordering that can not be deterministically modeled. 

When $h_\mu$ attains its maximum value of $1$, the stacking arrangement is as random as a toss of a coin. On the other extreme, if $h_\mu=0$, then the staking ordering is deterministically specified and the stacking order has infinite correlation, ideal long range order is achieved.

If the process has some amount of structure and some amount of randomness it will be useful to separate both quantities. Excess entropy is introduced to account for structure. The excess entropy is the mutual information between two halves of the stacking arrangement in the infinite length limit. For the Ising models dealt with here, the excess entropy will be given as the difference between the statistical complexity and the entropy density\footnote{Note that we are defining the Markov process as one where $\eta$ blocks are emitted instead of individual spins.\label{fn:indvs}} 
\begin{equation}
 E=C_{\mu}-h_{\mu}.
\end{equation}
For an ideally periodic process $E=\log D$ where $D$ is the period of the sequence.

Appendix \ref{sec:app} gives the expression that allows to calculate all probabilities involved in equations (\ref{eq:pr}), (\ref{eq:Cmu}) and (\ref{eq:entropyrateblock})

\begin{figure}
	\centering
	\includegraphics*[scale=0.7]{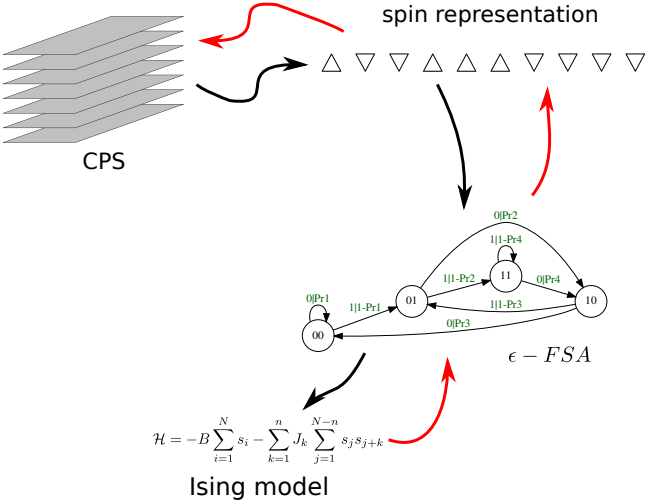}
	\caption{The analysis can be carried out in two directions. In one we start with the stacking arrangement and go to a deduction process that ends at discovering the Ising model that describes the system. In the second approach, an Ising model is built and from there, the $\epsilon$-machine is constructed that generates a binary sequences that statistically reproduces the stacking arrangement.}
	\label{fig:scheme}
\end{figure}

The analysis of the stacking process can be done in two different directions. In the first one, the stacking arrangement is known and coded into a binary array, from which an FSA is reconstructed and the underlying Ising model deduced. This is the inverse Ising problem. In the second direction an Ising model is constructed from where the $\epsilon$-machine is obtained. In turn, from the $\epsilon$-machine the binary code can be generated representing the stacking arrangement (Figure \ref{fig:scheme}); this is the analysis that will be done in what follows.

\section{Stacking sequence as a 1/2 next nearest neighbor Ising model in one dimension\label{sec:NNN}}

From the general expression (\ref{eq:isinghamiltonian}) consider the special case given by
\begin{equation}
 \mathcal{H}(s^N)=-J_{1}\sum\limits_{j=1}^{N-1}s_{j}s_{j+1}-J_{2}\sum\limits_{k=1}^{N-2}s_{k}s_{k+2}.\label{eq:nnnhamiltonian}
\end{equation}
The $\eta$ block is then 
\begin{equation}\eta_{dd}=\bigtriangledown\bigtriangledown, \eta_{du}=\bigtriangledown\bigtriangleup, \eta_{ud}=\bigtriangleup\bigtriangledown,\eta_{uu}=\bigtriangleup \bigtriangleup\label{eq:blocks}\end{equation}
we take the value of $1$ to represent spin $\bigtriangleup$, while $-1$ represents spin $\bigtriangledown$.

The transfer vector (\ref{eq:uv}) is given by
\begin{equation}
 |U \rangle=\left [ e^{-\frac{1}{2} \beta  (2 B-\text{J1})},e^{-\frac{\beta  \text{J1}}{2}},e^{-\frac{\beta  \text{J1}}{2}},e^{-\frac{1}{2} \beta  (-2 B-\text{J1})}\right ],
\end{equation}
and the transfer matrix (\ref{eq:vm}) is
\begin{equation}
 V=\left(
\begin{array}{cccc}
 e^{2 (-B+\text{J1}+\text{J2}) \beta } & e^{(\text{J1}-B) \beta } & e^{-(B+\text{J1}) \beta } & e^{-2 \text{J2} \beta } \\
 e^{-(B+\text{J1}) \beta } & e^{2 (\text{J2}-\text{J1}) \beta } & e^{-2 \text{J2} \beta } & e^{(B+\text{J1}) \beta } \\
 e^{(\text{J1}-B) \beta } & e^{-2 \text{J2} \beta } & e^{2 (\text{J2}-\text{J1}) \beta } & e^{(B-\text{J1}) \beta } \\
 e^{-2 \text{J2} \beta } & e^{(B-\text{J1}) \beta } & e^{(B+\text{J1}) \beta } & e^{2 (B+\text{J1}+\text{J2}) \beta } \\
\end{array}
\right)
\end{equation}

\begin{figure}
	\centering
	\includegraphics*[scale=0.7]{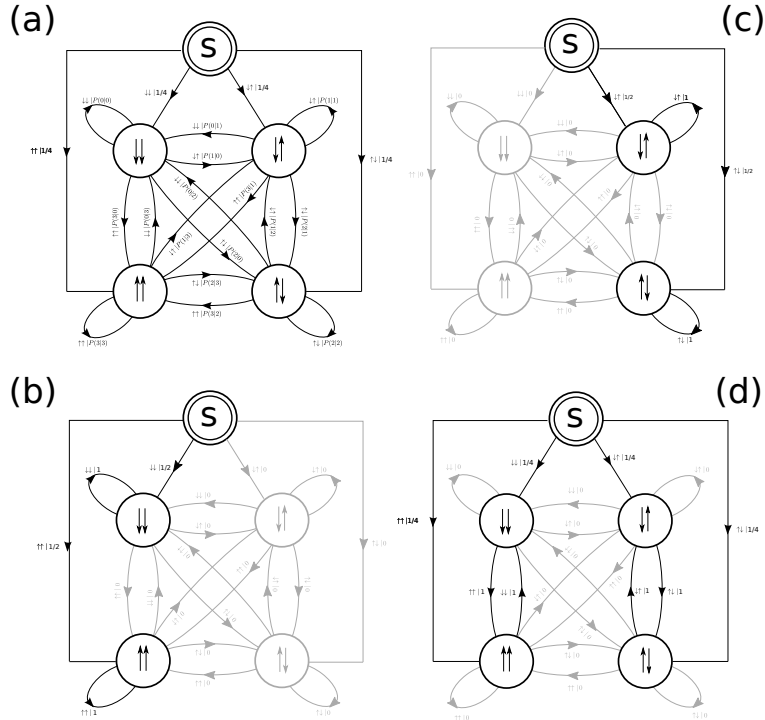}
	\caption{FSA for the NNNIM (a) The maximum connectivity FSA, (b) FCC configuration, (c) HCP configuration and (d) the DHCP configuration. Zero temperature ($\beta\rightarrow\infty$)}
	\label{fsm_nnn}
\end{figure}

Let us start by considering the basic state ($\beta\rightarrow\infty$). The range of interaction $n=2$ implies, at most, four possible causal states. The most general finite state automaton is shown in figure \ref{fsm_nnn}a. This FSA is not yet an $\epsilon$-machine, as  the stochastic matrix must be computed for different ratios of the interaction parameters $J2/J1$. A transient state $S$ is added to the FSA to account for the starting transition in the machine. Once the first transition is made, state $S$ becomes irrelevant as it is not visited anymore. In the stochastic matrix, the entries are ordered as $S, \eta_{dd}, \eta_{du}, \eta_{up}, \eta_{uu}$, as defined in (\ref{eq:blocks}).

For $J_2/J_1>-1/2$ and $J_1>0$, the stochastic matrix for the general FSA is given by 
\begin{equation}
 \displaystyle P=\left ( \begin{array}{ccccc}
			  0 & 1/2 & 0 & 0 & 1/2\\
                          0 & 1 & 0 & 0 & 0\\
                          0 & 0 & 0 & 0 & 0\\
                          0 &  0 & 0 & 0 & 0\\
                          0 & 0 & 0 & 0 & 1
                          \end{array}
\right ).
\end{equation}
Each entry $p_{ij}$ in the matrix gives the probability of making a transition from state $i$ to state $j$.
 
It follows that the states $\bigtriangledown \bigtriangleup$ and $\bigtriangleup \bigtriangledown$ are unreachable, and the only two recurrent states are $\bigtriangleup \bigtriangleup$ and $\bigtriangledown \bigtriangledown$.  They will be called the FCC states. Once one of these states is reached after the first step, the system stays on that state with certainty (Fig. \ref{fsm_nnn}b). Both recurrent states are equivalent, and therefore redundant. The corresponding $\epsilon$-machine is given by only one of the two FCC states and the stationary process is the one where the output symbol is always the same with total certainty. The statistical complexity is $C_{\mu}=0$ bits and the entropy density is $h_{\mu}=0$ bits/site.

For $J_2/J_1>1/2$ and $J_1<0$ the stochastic matrix for the general FSA is given by 
\begin{equation}
 \displaystyle P=\left ( \begin{array}{ccccc}
			  0 & 0 & 1/2 & 1/2 & 0\\
                          0 & 0 & 0 & 0 & 0\\
                          0 & 0 & 1 & 0 & 0 \\
                          0 & 0 & 0 & 1 & 0 \\
                          0 & 0 & 0 & 0 & 0
                          \end{array}
\right ).
\end{equation}
Now the states $\bigtriangleup \bigtriangleup$ and $\bigtriangledown \bigtriangledown$ are unreachable  and the states $\bigtriangledown \bigtriangleup$ and  $\bigtriangleup \bigtriangledown$ are both recurrent they will be called the HCP states. As before, once a recurrent state is reached, the system stays in that state with certainty (Fig. \ref{fsm_nnn}c).  Again, both recurrent states are equivalent and redundant. The corresponding $\epsilon$-machine is described by only one of the two recurrent states. The statistical complexity is $C_{\mu}=0$ bits and the entropy density is $h_{\mu}=0$ bits/site.

Finally, another region remains defined by $J_2/J_1<\frac{1}{2}$, $J_1<0$ and $J_2/J_1<-\frac{1}{2}$, $J_1>0$, where the stochastic matrix for the general FSA is given by 
\begin{equation}
 \displaystyle P=\left ( \begin{array}{ccccc}
			  0 & 1/4 & 1/4 & 1/4 & 1/4\\
                          0 & 0 & 0 & 0 & 1\\
                          0 & 0 & 0 & 1 & 0 \\
                          0 & 0 & 1 & 0 & 0 \\
                          0 & 1 & 0 & 0 & 0
                          \end{array}
\right ).
\end{equation}
This corresponds to alternating blocks of $\bigtriangleup \bigtriangleup$ and $\bigtriangledown \bigtriangledown$ (DHCP phase), which can be produced either by alternating between the FCC states $\bigtriangleup \bigtriangleup$ and $\bigtriangledown \bigtriangledown$, or between the HCP states $\bigtriangledown \bigtriangleup$ and $\bigtriangleup \bigtriangledown$ (Fig. \ref{fsm_nnn}d). Each pair of states defines the same dynamics and therefore they are redundant. The $\epsilon$-machine  is then described by any of them. The statistical complexity is $C_{\mu}=1$ bits and the entropy density is $h_{\mu}=0$ bits/site.

\subsection{The entropic analysis of the phase diagram.\label{sec:entropic}}

\begin{figure}
	\centering
	\includegraphics*[scale=1.8]{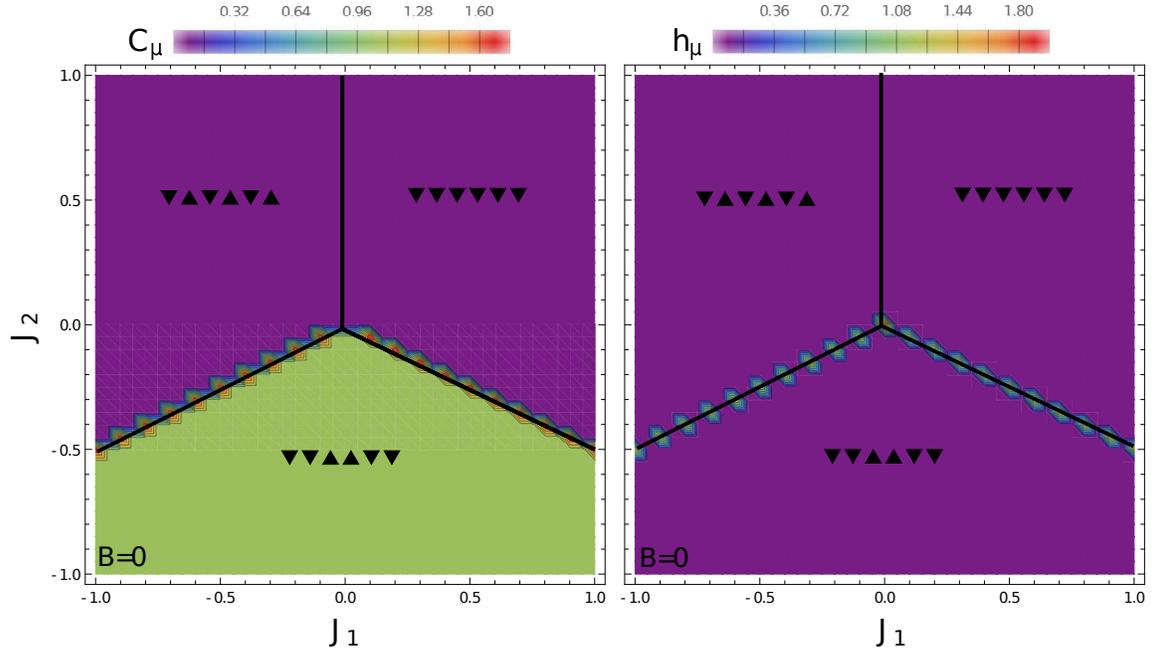}
	\caption{$C_\mu$ vs $J_1\times J2$ (left) and $h_\mu$ vs $J_1\times J2$ (right) for the fundamental states ($\beta \rightarrow \infty$).}
	\label{chbetainfb0}
\end{figure}

Figure \ref{chbetainfb0} shows the statistical complexity  and entropy density for interaction parameters in the range $J_1\times J_2$ within  $[-1,1]\times[-1,1]$ which reproduces a representative portion of the phase diagram.

\begin{figure}
	\centering
	\includegraphics*[scale=0.7]{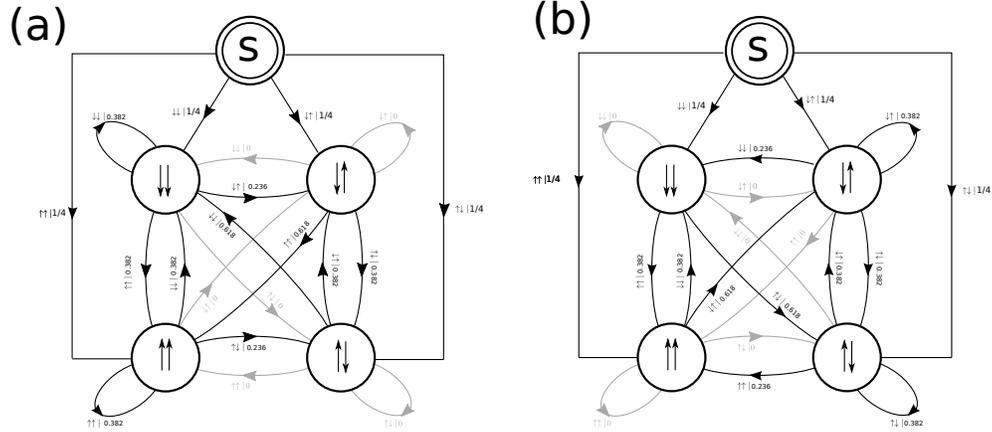}
	\caption{FSA at the boundaries between the (a) FCC and DHCP regions and, (b) the HCP and DHCP regions. Zero temperature ($\beta\rightarrow\infty$)}
	\label{fsm_nnn_border}
\end{figure}

The boundary lines $J_1=2J_2<0$  and  $J_1=-2J_2>0$ represent, respectively, the boundary between the DHCP and the HCP or FCC phases. In this boundary, the corresponding FSA is described by the four causal states, all strongly connected, which leads to high disorder ($h_{\mu}\approx 1.4$ bits/site) and high $C_{\mu}$ ($\approx 1.85$ bits). Figure \ref{fsm_nnn_border}a is the FSA describing the FCC-DHCP boundary. It can be seen that the FSA is a mix of the FCC FSA and the DHCP FSA. There is a strong connection (probability $=0.618$) between the FCC state pair and the HCP state pair, while the FCC states are self referenced with a non-negligible probability ($=0.382$). As a result of this interconnection, besides the FCC and DHCP phases, polytypes with larger periodicity have probability of occurrence above zero. 

Following \cite{estevez08} all polytypes up to length 12 were generated and the probability of their occurrence was calculated using the $\epsilon$-machine description. The sequence that represents energetically stable  polytypes are represented by close loops in the $\epsilon$-machine. Those that are not close loops are the result of the antiphase-boundary between two phases, which can be the same or not. Such frontiers are metastable. The analysis shows an important advantage of the $\epsilon$-machine description over other approaches, it allows discovering, instead of recognizing, polytypes by identifying all possible close loops. Table \ref{tbl:fccdhcp} shows all stable polytypes appearing at the phase boundary, as well as the metastable sequences, appearing as antiphase frontier. 

A similar dynamics can be found in the HCP-DHCP boundary (Figure \ref{fsm_nnn_border}b). Now the HCP states are the self-referenced ones. Connectivity between the FCC and HCP states is equally strong. Table \ref{tbl:hcpdhcp} shows the polytypes up to a length of $12$. At this boundary, using Zhdanov code, the orderings $1122$, $111122$, $11111122$, $112222$, $1111111122$ and $11112222$ are the result of adjacent HCP and DHCP blocks. 

The nature of the boundary between the FCC and HCP phases ($J_1=0, J_2>0$) is different. The FSA representation of the boundary shows that once the system gets into the FCC or the HCP ordering, it will stay there. Within the sequential model used disorder, therefore, is an ensemble property (not of a single stacking) and it will be the result of the initial probability of the two ordering, then, for a single stacking, $h_\mu=0$. \footnote{If syntactic coalescence \cite{triguniyat91} is taken into account then  interspersion of FCC and HCP regions should happen resulting in $h_\mu=1$. In other cases, when a strongly connected digraph represents the dynamics of the system, syntactic coalescence does not change the analysis. } 

\begin{figure}
	\centering
	\includegraphics*[scale=1.8]{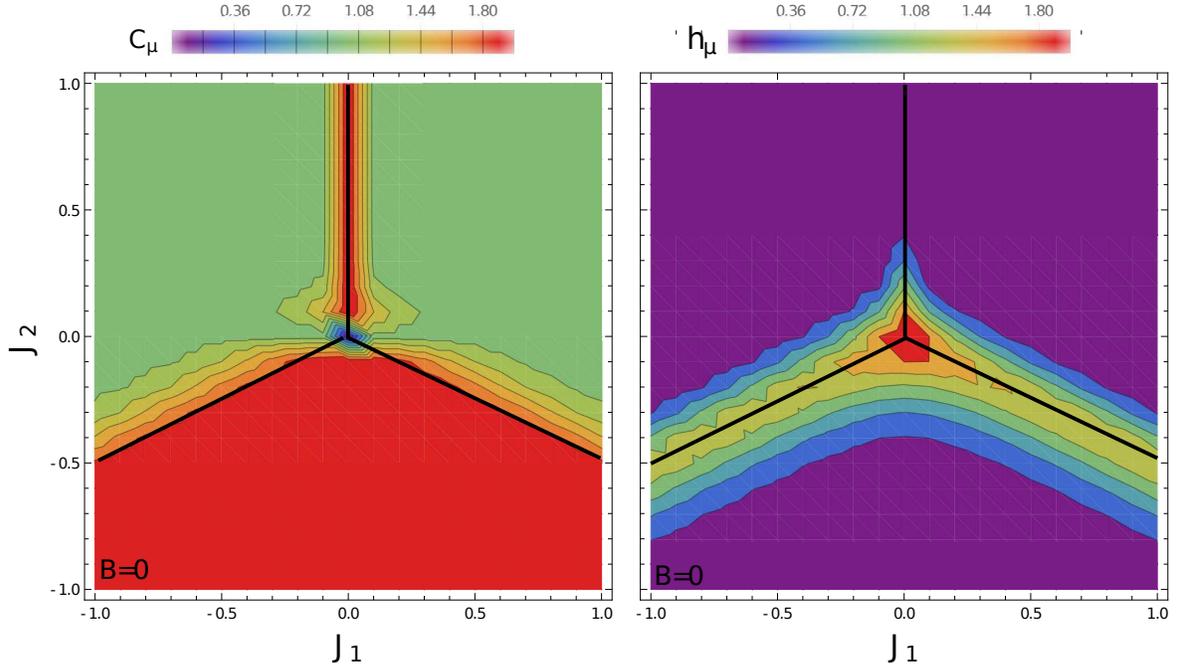}
	\caption{$C_\mu$ vs $J_1\times J2$ (left) and $h_\mu$ vs $J_1\times J2$ (right) with temperature above zero ($\beta =5$).}
	\label{chbeta5b0}
\end{figure}

As soon as temperature is increased above zero, the systems lose order. This in turn implies the increase of the number of causal states in each region. The FSA (Figure \ref{fsm_nnn}a) describes the dynamics of the $\epsilon$-machine; as a result $C_\mu$ increases in all regions (Figure \ref{chbeta5b0} left), the system needs more resources to predict its dynamics and $h_\mu$ (Figure \ref{chbeta5b0} right) increases as the balance between $J_2$ and $J_1$ draws the system near a phase boundary. 

Consider $\beta=5$. In all regions the full $\epsilon$-machine is strictly needed, yet deep in the FCC region (we took $J_2=J_1=0.2$ for the numerical values that follow), the probability of being in the $\bigtriangleup \bigtriangleup$ or $\bigtriangledown \bigtriangledown$ states is almost one half for both ($=0.498$), while for the other two HCP states are near zero ($=0.00164$). While at zero temperature, once one of the two FCC states was reached the system did not leaved that state, thermal disorder, no matter how small, weakly connects the two FCC states making them equally probable. This is the $\epsilon$-machine realization of the zero magnetization state, known for the one-dimensional Ising model\footnote{Magnetization for our purposes can be defined as the sum of all spins making the stacking arrangement, when $1$ is taken to represent spin $\bigtriangleup$, while $-1$ represents spin $\bigtriangledown$.} \cite{greiner95}. 

The transition between the two FCC states can be done directly $\cdots \bigtriangleup \bigtriangleup \bigtriangleup \bigtriangleup \bigtriangledown \bigtriangledown \bigtriangledown \bigtriangledown \cdots$ with probability $0.00246$, giving a two spin $\bigtriangleup \bigtriangledown$ slab at the boundary, or it can be mediated by one of the HCP states $\cdots \bigtriangleup \bigtriangleup \bigtriangleup \bigtriangledown \bigtriangleup \bigtriangledown \bigtriangledown \bigtriangledown$ with almost the same probability $0.00245$. In the last case, a four spin HCP boundary $\bigtriangleup \bigtriangledown \bigtriangleup \bigtriangledown$ separates the two FCC regions. Summarizing, the FCC state is now described by long sequences of parallel spins (domains), separated with equal probability by two and four spin HCP boundaries. 

In the HCP region (we took $J_2=-J_1=0.2$ for the numerical values that follow), the states $\bigtriangledown \bigtriangleup$ and $\bigtriangleup \bigtriangledown$ are the ones with equal and higher probability of $0.498$, while the FCC states have a much smaller probability of $0.0016$. Again magnetization is zero, both HCP states are weakly coupled  by a two spin or four spin FCC boundary, each equally probable with probability $0.00246$.

In the DHCP region  (we took $J_2=-0.5$ and $ J_1=-0.2$ for the numerical values that follow) all four states are equally probable, with the FCC states with probability $0.248$, slightly smaller probability than the HCP states with probability $0.252$. At zero temperature the two FCC states were decoupled from the two HCP states, now, at $\beta=5$ there is a weak coupling, given by a probability $0.0359$ of making a transition between both pair of states. The output are long sequences of alternating $\bigtriangleup \bigtriangleup$ and $\bigtriangledown \bigtriangledown$ blocks, where eventually, one spin boundaries can be found (three spin FCC block) 
\[
\cdots \bigtriangleup \bigtriangleup \bigtriangledown \bigtriangledown \bigtriangleup \bigtriangleup \circ \underline{\bigtriangledown \circ \bigtriangledown  \bigtriangledown } \bigtriangleup \bigtriangleup \bigtriangledown \bigtriangledown \cdots
\]
where the $\circ$ signals the boundary.

On the light of this analysis, some of the long period polytypes reported by \cite{price84} for temperatures above zero,  are boundary polytypes between two blocks of $3C$, $2H$ or $4H$ phases stabilized by entropic factors. This is certainly the case for the reported $33$ ($6H$) polytype and the $12$ sequence, which should happen at the antiphase boundary between $3C$ blocks; or the different $2^p3$ sequence (the notation used by  \cite{price84} is followed here, which is straightforward to understand) that should happen when a $4H$ and $3C$ blocks meet. Also, the reported sequences of the type $12^p$ should occur at the boundary between $2H$ and $4H$ blocks. The polytypes $44$ and $55$, which Price and Yeomans did not observed in their ANNNI model, have been found at the FCC-DHCP border.

\begin{figure}
	\centering
	\includegraphics*[scale=1]{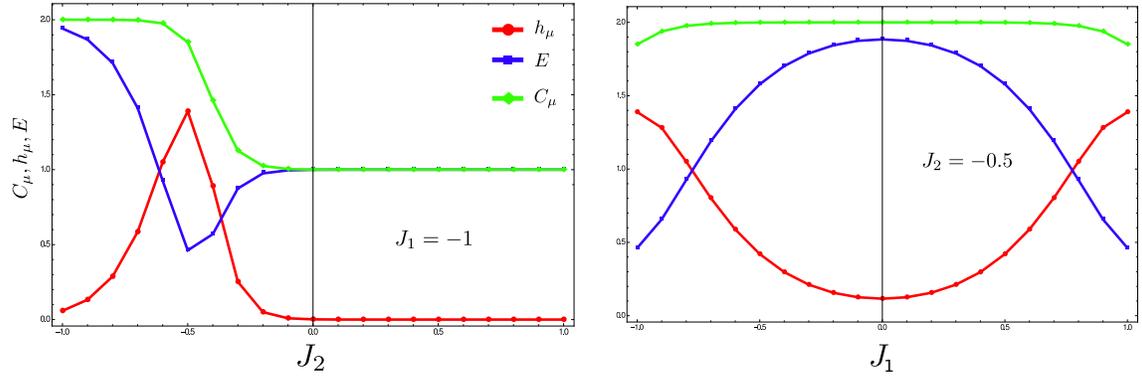}
	\caption{Statistical complexity ($C_\mu$), entropy density ($h_\mu$) and excess entropy as a function of $J_2$ for (left) $J_1=-1$ and (right) as a function of $J_1$ for $J_2=-0.5$.}
	\label{complj2}
\end{figure}

Figure \ref{complj2}-left shows the evolution of $C_\mu$, $h_\mu$ and excess entropy as $J_2$ increases, while keeping $J_1$ fixed. To the left, a transition from the DHCP to the HCP region is seen as a significant increase of disorder $h_\mu$ reaching its maximum at the boundary, while the statistical complexity goes from a four state dynamics to an essentially two (HCP) state dynamics, and therefore dropping from very near $2$ bits to $1$ bit. Excess entropy $E$, on the other hand, has a minimum at the phase boundary, where much of the $C_\mu$ can be accounted as non correlated disorder. 

Consider now the evolution of $C_\mu$, $h_\mu$ and excess entropy as $J_1$ increases, while keeping $J_2$ fixed at $-0.5$. Looking at Figure \ref{chbeta5b0} it can be seen that the starting point is at the boundary of the HCP-DHCP phases, going through the DHCP phase and ending at the boundary between this last phase and the FCC one. In figure \ref{complj2}-right, the statistical complexity $C_\mu$ stays almost constant in the whole range of $J_1$values. Entropy density $h_\mu$ starts and ends at a symmetrical maximum corresponding to the boundaries, while it decreases from there, reaching a minimum at $J_1=0$. Disorder decreases more the ``deeper'' the system gets into the DHCP phase. Excess entropy behavior is then result of a diminishing entropy density over a nearly constant $C_\mu$ and it reaches a maximum at the lowest $h_\mu$, with $J_1=0$. The deeper into the DHCP region, the more structured the system dynamic becomes. 

\begin{figure}
	\centering
	\includegraphics*[scale=2.0]{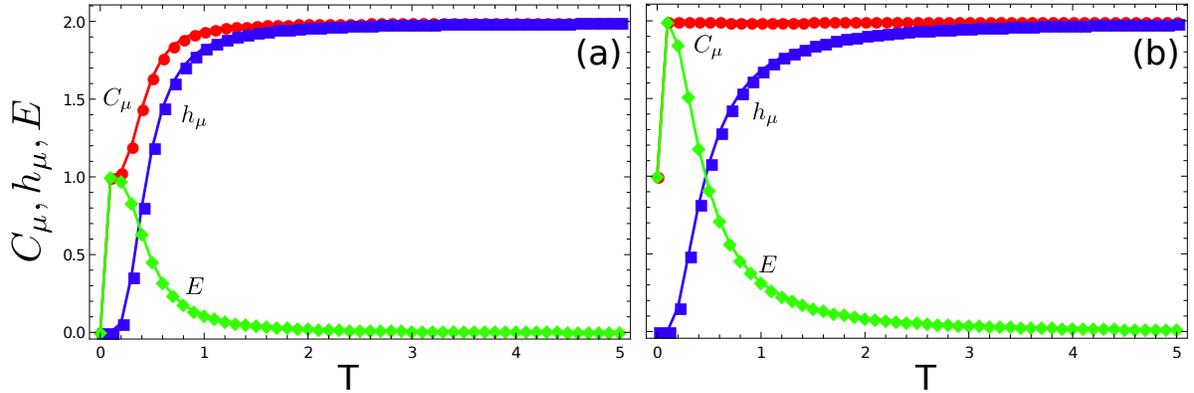}
	\caption{$C_\mu$, $E_\mu$ y $h_\mu$ as a function of temperature $T=1/\beta$ for $B=0$. $a)$ FCC regime $J_1=J2=0.2$. $b)$ DHCP dynamic $J_1=-0.2,J2=-0.5$.}
	\label{vsT}
\end{figure}

Finally, the dynamics as a function of temperature has been explored. In the FCC case (Figure \ref{vsT}a) entropy density increases with temperature as expected. As soon as $T\neq 0$ a jump in the excess entropy signals that the system has vanishing magnetization at zero applied field. After that first jump, $E$ decreases monotonically as temperature increases, witnessing the loss of correlation in the system. Statistical complexity $C_\mu$ is zero at $T=0$ and the perfect FCC case has only one causal state. As soon as thermal agitation gets in, the system becomes a two state dynamic (both FCC states become connected) and $C_\mu$ jumps to a value of one. Further increasing the temperature (but keeping it finite valued $T<\infty$) results in all four states coming into play and $C_\mu$ increases as the system becomes more disordered and the distribution of probability over the causal states becomes more uniform. The same explanation is valid for the DHCP region (Figure \ref{vsT}b). The difference is that the initial jump in $E$ is from a two-state dynamic at $T=0$ to a four-state dynamic at $T\neq 0$ ($E=2$ bits). From there on, it decreases as temperature increases. $C_\mu$  keeps constant with temperature after the first jump, as the four states are evenly involved as soon as thermal noise is a factor in the system. For $T\rightarrow \infty$, rows in the stochastic matrix become identical, the $\epsilon$-machines collapse to a single state with maximum entropy and zero excess entropy (statistical complexity also becomes zero) equivalent to the toss of a coin.

\section{Discussion and conclusion\label{sec:discussion}}

The Hamiltonian given by equation (\ref{eq:nnnhamiltonian}) is not the most general one when dealing with polytypism. \cite{shaw90}, following \cite{cheng88}, have reported a more general Hamiltonian over spins, as coded by the  H\"{a}gg (Nabarro-Frank) code. It is given by the sum of pairwise terms as well as the sum of terms made by monoids of an even number of spins. 
\begin{equation}
\begin{array}{rl}
 \mathcal{H}(s^L)=&-B \sum\limits_{i=1}^{N}s_{i}-J_{1}\sum\limits_{j=1}^{N-1}s_{j}s_{j+1}-J_{2}\sum\limits_{k=1}^{N-2}s_{k}s_{k+2}\\\\&-J_{3}\sum\limits_{k=1}^{N-3}s_{k}s_{k+3}-K\sum\limits_{k=1}^{N-3}s_{k}s_{k+1}s_{k+2}s_{k+3}.
\end{array}
 \label{eq:chenghamiltonian}
\end{equation}
Terms made by an odd number of monoids are excluded as they will not remain invariant under a $s_i \rightarrow -s_i$ transformation, which is a necessary symmetry condition (it represents a $180^o$ rotation of the whole stacking arrangement). Monoid terms above pairwise interaction are introduced for several reasons. One of them accounts for difficulties in justifying interactions between fictitious entities (pairs of layers), as coded by the  H\"{a}gg procedure. Hamiltonian such as equation (\ref{eq:chenghamiltonian}) attempts to deal with such difficulties by taking into account the intermediate spins (last term in the equation), which should eliminate geometric and symmetry inconsistencies. The reader should refer to the paper by \cite{shaw90} for a more thorough discussion. 

For SiC, the values of the $J_1$ and $J_2$ parameters, as given by \cite{cheng88}, are of the same order of magnitude and one  order larger than the $J_3$ and $K$ parameters. Yet, in the case of ZnS, as calculated by \cite{engels90}, $J_1$ is in the order of $10^{-3}$ eV, while $J_2$ is two orders below, being $J_3$ and $K$ negligible. In such case the use of Hamiltonians of the type given by (\ref{eq:nnnhamiltonian}) with $B=0$ is justified. For ZnS, the ratio $J_2/J_1=-4.28\times 10^{-2}>-1/2$, with $J_2<0$, corresponds to the FCC region. The appearance of different order polytypes is then argued in terms of the phonon contribution at $T>0$, which gives a dependence of the interaction parameters $J_i$ on temperature. 

The same explanation of the temperature dependence of the interaction parameters is used to account for the dependence of phase transformation on temperature as well, while still relying on the validity of a one dimensional Ising model where phase transformations should not happen. If $J_i$ is taken as a function of temperature, pressure or composition, then the dynamics of disorder and polytypic transformation would be described by the actual path in any ``structural parameter`` vs $J_1\times J_2$ diagram, such as that of Figure \ref{chbeta5b0}. The diagram itself changes with the temperature. From the discussion in the previous section, specially about the plot in Figure \ref{vsT}, the effect of raising temperature is essentially to increase the entropy density of the system, while, consequently, the amount of ''structured'' stacking decreases.  Disorder creeps in as bands or region along the boundaries, which can be followed by looking at $h_\mu$ (see Figure \ref{chbeta5b0}b). Calculations show that such bands are of increasing width with growing temperature. The boundaries involving the DHCP phase are in general more disordered than the HCP-FCC boundary. For other changing external factors (pressure, composition, etc), the phase diagram does not change.

ZnS exhibits a large number of polytypic variants. In consequence, if one is to assume a Hamiltonian of the type given by  (\ref{eq:nnnhamiltonian}) as valid and that ZnS polytypes are of thermodynamic origin; then  temperature drives the system into the disorder band around the boundaries involving the DHCP region. As reported by \cite{engels90}, the $FCC$, $22$, $33$ polytypes are commonly observed in experiments. Looking at Table \ref{tbl:fccdhcp}, the occurrence of the $6H$ polytype seems to suggest that the FCC-DHCP disordered band is more likely to be approached with temperature starting at the FCC structure, which as already discussed, is the stable ground state. 

Let us consider the Cobalt-Nickel alloy. The picture in the phase transformation of this alloy is completely different. In the CoNi phase transformation between HCP and FCC structure occurs at temperatures depending on the amount of Ni. In the case of metal alloys it is necessary to consider long range order, which can be taken asymptotically as an oscillating damping term \cite{blandin68}.  \cite{bruinsma85} have shown that the Hamiltonian taking into account this long range term can give rise, at special tuning of the parameters, to a wide number of polytypes depending on the electron concentration. This seems not to be the case of the CoNi alloy. We have performed X-ray diffraction experiments in Co-Ni alloys of different compositions. Although detailed results will be reported elsewhere, it is appropriate to mention that, in certain regions of composition, extensive disorder occurs at a wide range of temperature. Yet, no high order polytype could be identified. Besides, disorder frozen within each phase and remanent of the other phase is found even well above or below the transformation temperature. 

The mineral wollastonite ($CaSiO_3$) exhibits two polytipic variations: one FCC and the other DHCP. \cite{price84} mention experiments performed by others suggesting that at high temperatures DHCP ordering stabilizes relative to the FCC structure. Moreover, for a temperature range between those where the FCC and the DHCP phases are dominant, other polytypes  with structure $33$, $44$ and $55$ were also found. While \cite{price84} consider the $44$ and $55$ ordering metastable, our results proves that they can appear as stable phases, as can be seen in Table \ref{tbl:fccdhcp}.

If the model used in this work is taken as a qualitative picture, one can understand the lack of higher order polytypes. Their non-occurrence is the result of a phase transition resulting from a path starting at the HCP phase and ending at the FCC phase, while going through their common boundary. In the experiment it was found that the coherence length, as a measure of the disorder density, abruptly climbs when reaching the reported phase transition for pure Cobalt, which could be understood as the entering into de disorder band near the HCP-FCC border. 

The $\epsilon$-machine description of the stacking ordering, even within this very short range Ising model, allows for the occurrence of stable polytypes as well as metastable sequences. In this way, some experimental reports of long period polytypes, specially those done by electron microscopy or other local probe techniques, could be the result of ``spotting'' a metastable sequence, which does not necessarily amount to the happening of a given polytype. Care must be taken to distinguish between the above chance of occurrence of a stacking sequence and the true happening of a polytype structure.

In conclusion, we have developed the computational mechanics of a next nearest neighbor interaction model applied to the stacking ordering problem in close packed structure. In doing so, the general procedure for calculating the stochastic matrix from the Gibbs field is reported. The building of the $\epsilon$-machine has allowed a thorough discussion of disorder and polytypism within the model. $\epsilon$-machine allows to quantify disorder through the use of the entropy density $h_\mu$, and ``structure'' through the use of excess entropy $E$, something that other approaches are unable to do. In this sense, the use of computational mechanics applied to the stacking problem, as develop by Varn, Crutchfield and coworkers, goes beyond the possible reconstruction of the stacking dynamics from the experiment. It allows a powerful approach to the origins of polytypism in solids as well, an open new venue into the understanding of an old problem.

 \appendix
 \section{The Ising model as a Markov process}\label{sec:app}

According to the Perron-Frobenius theorem for a positive square matrix $V$, the following expression holds
\[
  \displaystyle \lim_{L\rightarrow \infty} \frac{V^L}{\lambda_0^L}=|_ra_{0}\rangle \langle _la_0|
\]
where $\lambda_{0}$ is the non-degenerate largest eigenvalue of $V$, $_{r}a_{0}$ ($_{l}a_{0}$) is the right (left) eigenvector associated to $\lambda_{0}$. The normalization condition $\langle _la_0|_ra_0\rangle=1$ is taken.

For $L\gg 1$ the partition function given by (\ref{eq:partition})
\[
 Z_{N}=\langle U | V^{L-1} | U \rangle
\]
can then be written as
\[
 Z_{N}=\left ( \langle U| _ra_0\rangle \langle _la_0|U\rangle \right ) \lambda_0^{L-1}=M \lambda_0^{L-1}.
\]
 
From the partition function, the probability of a given configuration $S^{N}$ of length $N=Ln$, will be
\begin{equation}
\begin{array}{ll}
 Pr(S^{N})&=\frac{1}{Z_{N}}\exp \left [ -\beta H(S^{N})\right ]\\ \\
 &=\frac{U_{\eta_{1}} U_{\eta_{L-1}}}{M \lambda_{0}^{L-1}}\prod\limits_{i=1}^{L-1}V_{\eta_{i}\eta_{i+1}}.
 \end{array}\label{eq:probsl}
\end{equation}

The probability of the $j$ block having value $\eta_{j}$, while the rest of the spin configuration $s^{N}-\eta_{j}$ (that is the spin configuration $S^N $not considering $\eta_j$) is fixed, will be given by the multiplication probability rule
\begin{equation}
 Pr(\eta_{j}| s^{N}-\eta_{j})=\frac{Pr(S^{N})}{Pr(S^{N}-\eta_{j})}.\label{eq:probrule}
\end{equation}
Now, according to equation (\ref{eq:probsl}),
\begin{equation}
\begin{array}{rl}
 \displaystyle Pr(s^{N}-\eta_{j})& =\left( \frac{U_{\eta_{1}} U_{\eta_{L-1}}}{M \lambda_{0}^{L-1}} \prod\limits_{\mathclap{\substack{i=1\\ i\neq j, j-1}}}^{L-1} V_{\eta_{i}\eta_{i+1}}\right)\\\\&U_{\eta_{j-1}}U_{\eta_{j+1}}\sum\limits\limits_{\{\eta_{k}\}} V_{\eta_{j-1}\eta_{k}}V_{\eta_{k}\eta_{j+1}}\label{eq:pretasineta}.
 \end{array}
\end{equation}
The sum in the second term is over all possible $\eta$ blocks. Substituting (\ref{eq:probsl}) and (\ref{eq:pretasineta}) into (\ref{eq:probrule}) we get
\begin{equation}
 Pr(\eta_{j}| s^{N}-\eta_{j})=\frac{V_{\eta_{j-1}\eta_{j}}V_{\eta_{j}\eta_{j+1}}}{\sum\limits_{\{\eta_{k}\}} V_{\eta_{j-1}\eta_{k}}V_{\eta_{k}\eta_{j+1}}},\label{eq:localcharact}
\end{equation}
valid for blocks inside the sequence and
\begin{equation}
 Pr(\eta_{1}| s^{N}-\eta_{1})=\frac{U_{\eta_1}V_{\eta_{1}\eta_{1}}}{\sum\limits_{\{\eta_{k}\}} U_{\eta_{k}}V_{\eta_{k}\eta_{1}}},\label{eq:localcharact1}
\end{equation}
for the first block, similar expression can be written for the last block. 

From this last equation it is clear that
\begin{equation}
 Pr(\eta_{j}| s^{N}-\eta_{j})=Pr(\eta_{j}|\eta_{j-1},\eta_{j+1})
\end{equation}
$Pr(\eta_{j}|\eta_{j-1},\eta_{j+1})$ is called the local characteristic and the whole system is termed a Markov (Gibbs) random field.

Using Bayes theorem the local characteristic can be written in terms of the stochastic matrix entries (\ref{eq:hmm})
\begin{equation}
 \begin{array}{l}
  Pr(\eta_{i}, \eta_{i-1},\eta_{i+1})=Pr(\eta_{i}| \eta_{i-1},\eta_{i+1})Pr(\eta_{i+1},\eta_{i-1})=\\\\
=Pr(\eta_{i}| \eta_{i-1},\eta_{i+1})Pr(\eta_{i+1}|\eta_{i-1})Pr(\eta_{i-1}),
 \end{array}\label{eq:protra}
\end{equation}
from where
\begin{equation}
 \begin{array}{l}
  Pr(\eta_{i+1}|\eta_{i-1})=\frac{Pr(\eta_{i}, \eta_{i-1},\eta_{i+1})}{Pr(\eta_{i}| \eta_{i-1},\eta_{i+1})Pr(\eta_{i-1})}.
 \end{array}\label{eq:protrb}
\end{equation}

\subsection{Calculating the stochastic matrix.\label{sec:sm}}

To calculate the stochastic matrix, a functional approach will be followed.

Consider two possible blocks $\eta_i$ and $\eta_j$, the corresponding entry in the stochastic matrix $Pr(\eta_i|\eta_{j})$ can be calculated through
\begin{equation}
\displaystyle Pr(\eta_i|\eta_j)=\frac{Pr(\eta_i,\eta_{j})}{Pr(\eta_{j})},\label{eq:sm1}
\end{equation}
where $Pr(\eta_i,\eta_{j})$ is the probability of finding a sequence $\eta_j\, \eta_{i}$. Taking into account that 
\[Pr(\eta_{j})=\sum\limits_{\{\eta_i\}} Pr(\eta_i,\eta_{j}),\]
where the sum is over all possible blocks $\eta_{i}$, equation (\ref{eq:sm1}) can be written as a function only of  $Pr(\eta_i,\eta_{j})$
\begin{equation}
\displaystyle 	Pr(\eta_i|\eta_{j})=\frac{Pr(\eta_i,\eta_{j})}{\sum\limits_{\eta_i}Pr(\eta_i,\eta_{j})}\label{eq:preiek}
\end{equation}

For an arbitrary function $q(\eta_i,\eta_{j})$ acting over spin blocks, consider the functional
\begin{equation}
\displaystyle Z[q(\eta_i,\eta_{j})]=\sum\limits_{\{S^N\}} \exp\left \{-\beta\left[ \mathcal{H}(S^N)-\sum\limits_i q(\eta_i,\eta_{j})\right]\right \} 
\end{equation}
The sum is performed over all possible spin configurations. As before, $\mathcal{H}(s^N)$ is the energy of the system. It is clear that $Z(0)=Z[h(\eta_i,\eta_{j})=0]$ is the partition function of the system given by (\ref{eq:part}). Take the derivative of $\ln Z[q(\eta_i,\eta_{j})]$ with respect to $q(\eta_k, \eta_{p})$, and making $q(\eta_i,\eta_{j})=0$ for all $\eta_i$ results in 

\begin{equation}
\begin{array}{rl}
\frac{\partial \ln Z[q(\eta_i,\eta_{j})]}{\partial q(\eta_k,\eta_{p})}|_{q=0} &=\frac{\beta L}{Z(0)}\sum\limits_{\{s^N\}}\delta(\eta_i,\eta_k)\delta(\eta_{j},\eta_{p}) e^{-\beta \mathcal{H}(s^N)}\\\\&=\beta L Pr(\eta_k,\eta_{p})
\end{array}\label{eq:z1}
\end{equation}
where $L$ is the number of blocks in the sequence ($N=Ln$) and $\delta(x,y)$ is the Kronecker delta over $x$ and $y$, which gives 1 if $x=y$ and zero otherwise.

If $\lambda_0'$ is the dominant eigenvalue of the transfer matrix built for the Hamiltonian $\mathcal{H}'=\mathcal{H}(s^L)-\sum\limits_i q(\eta_i,\eta_{j})$ , then $Z[q(\eta_i,\eta_{j})]=M'{\lambda_0'}^L$ for $L\gg 1$ and
\begin{equation}
\frac{\partial \ln Z[q(\eta_i,\eta_{j})]}{\partial q(\eta_k,\eta_{p})}|_{q=0} =L(\frac{1}{\lambda_0'} \frac{\partial \lambda_0'}{\partial q(\eta_k,\eta_{p})})|_{q=0}\label{eq:z2}
\end{equation}

From equation (\ref{eq:z1}) and (\ref{eq:z2}) one finally gets
\begin{equation}
Pr(\eta_k,\eta_{k-1})=\frac{1}{\beta}(\frac{1}{\lambda_0'} \frac{\partial \lambda_0'}{\partial q(\eta_k,\eta_{k-1})})|_{q=0}\label{eq:thesmexp}
\end{equation}

Expression (\ref{eq:thesmexp}) is general for deriving the stochastic matrix from the transfer matrix in the Ising model. 

Summarizing, in order to calculate the stochastic matrix, one proceeds as follow:
\begin{enumerate}
\item Build the transfer matrix $V'$ for the Hamiltonian $\mathcal{H}'=\mathcal{H}(s^L)-\sum\limits_i q(\eta_i,\eta_{i-1})$, where $q(\eta_i,\eta_{i-1})$ is an arbitrary function.
\item Calculate the dominant eigenvalue $\lambda_{0}^{'}$ for $V'$.
\item Calculate $Pr(\eta_i,\eta_{i-1})$ using expression (\ref{eq:thesmexp}).
\item Calculate $P(\eta_i|\eta_{i-1})$ using expression (\ref{eq:preiek}).
\end{enumerate}

\section{Acknowledgment}
 This work was partially financed by FAPEMIG under the project BPV-00047-13 and computational infrastructure support under project APQ-02256-12. EER wishes to thank AvH  for a fellowship renewal grant and the financial support under the PVE/CAPES grant 1149-14-8. Elettra project ­ 20115278 and 20110034 are acknowledged for experimental support, as well as ICTP financial support. RLS wants to thank the support of CNPq through the projects 309647/2012-6 and 304649/2013-9.


\newpage

\begin{table}
\caption{Some of the possible polytypes appearing at the FCC-DHCP border at $T=0$($\beta=\infty$). Sequence where determined only for close loops along the FSA (same starting and ending state) except for those marked with an arrow. L: The sequence length, Probability: probability of finding the sequence in the stacking arrangement, Ramsdell: Ramsdell notation of the corresponding polytype, Zhdanov: Zhdanov symbol of the corresponding polytype, Sequence: The sequence in  Nabarro-Frank notation}
\centering
\begin{tabular}{rllllr}
	L && Probability & Ramsdell & Zhdanov &  Sequence \\
	\hline\\
	
	 3 &&    0.447   &  3C  &  $\infty$  &  $\bigtriangleup\bigtriangleup \bigtriangleup$\\
	 4 &&  0.382 &  4H  &  $22$     &  $ \bigtriangleup \bigtriangleup \bigtriangledown \bigtriangledown$\\
	$\rightarrow$5 &&  0.21 & 5R & 14  &  $\bigtriangleup  \bigtriangledown \bigtriangledown \bigtriangledown\bigtriangledown$\\
	6 && 0.236 &  6H  &  $33$ & $\bigtriangleup\bigtriangleup\bigtriangleup\bigtriangledown\bigtriangledown  \bigtriangledown$\\
	7&& 0.367 & 7R  &  $25$  & $\bigtriangleup\bigtriangleup\bigtriangledown\bigtriangledown  \bigtriangledown\bigtriangledown\bigtriangledown$\\
	8 && 0.130  & 8H  & $44$ &$\bigtriangleup\bigtriangleup\bigtriangleup\bigtriangleup\bigtriangledown\bigtriangledown \bigtriangledown \bigtriangledown$ \\
	 9 && 0.186   & 9R  & $36$  & $\bigtriangleup\bigtriangleup\bigtriangleup\bigtriangledown\bigtriangledown\bigtriangledown \bigtriangledown \bigtriangledown\bigtriangledown$ \\
	$\rightarrow$9 && 0.062   & 9R  & $1224$  & $\bigtriangleup\bigtriangledown\bigtriangledown\bigtriangleup\bigtriangleup\bigtriangledown \bigtriangledown \bigtriangledown\bigtriangledown$ \\
	 10 && 0.133   & 10R  & $28$  &$\bigtriangleup\bigtriangleup\bigtriangledown \bigtriangledown \bigtriangledown\bigtriangledown\bigtriangledown \bigtriangledown \bigtriangledown\bigtriangledown$ \\
	 10 &&  0.111   & 10H  & $2233$  & $\bigtriangleup\bigtriangleup\bigtriangledown \bigtriangledown \bigtriangleup\bigtriangleup\bigtriangleup  \bigtriangledown \bigtriangledown\bigtriangledown$\\
	 10 && 0.065   &  10H  & $55$ &$\bigtriangleup\bigtriangleup\bigtriangleup\bigtriangleup\bigtriangleup\bigtriangledown\bigtriangledown\bigtriangledown \bigtriangledown \bigtriangledown$ \\
	 11 &&  0.09 & 11R  & $47$ & $\bigtriangleup\bigtriangleup\bigtriangleup\bigtriangleup\bigtriangledown\bigtriangledown\bigtriangledown\bigtriangledown \bigtriangledown \bigtriangledown\bigtriangledown$ \\
	 11 && 0.078 & 11R  &$2324$  & $\bigtriangleup\bigtriangleup\bigtriangledown\bigtriangledown\bigtriangledown\bigtriangleup\bigtriangleup\bigtriangledown \bigtriangledown \bigtriangledown\bigtriangledown$\\
	 11 &&  0.08 & 11R  & $2225$  & $\bigtriangleup\bigtriangleup\bigtriangledown\bigtriangledown\bigtriangleup\bigtriangleup\bigtriangledown\bigtriangledown \bigtriangledown \bigtriangledown\bigtriangledown$ \\
	 12 && 0.061 & 12R  & $39$ & $\bigtriangleup\bigtriangleup\bigtriangleup\bigtriangledown\bigtriangledown\bigtriangledown\bigtriangledown\bigtriangledown \bigtriangledown \bigtriangledown\bigtriangledown\bigtriangledown$ \\
	 12 && 0.053 & 12R  & $2343$  &$\bigtriangleup\bigtriangleup\bigtriangledown\bigtriangledown\bigtriangledown\bigtriangleup\bigtriangleup\bigtriangleup \bigtriangleup \bigtriangledown\bigtriangledown\bigtriangledown$\\
	 12 && 0.054 &  12H  &  $2244$  &  $\bigtriangleup\bigtriangleup\bigtriangledown\bigtriangledown\bigtriangleup\bigtriangleup\bigtriangleup\bigtriangleup \bigtriangledown \bigtriangledown\bigtriangledown\bigtriangledown$ \\
\end{tabular}\label{tbl:fccdhcp}
\end{table}

\begin{table}
\caption{Some of the possible polytypes appearing at the HCP-DHCP border at $T=0$($\beta=\infty$). Conditions and notation follows Table 1.}
\centering
\begin{tabular}{rlllll}
		L && Prob. & Ramsdell & Zhdanov &  seq. \\
		\hline\\
		 2 &&    0.276   &  2H  &  $ 11 $  & $\bigtriangleup\bigtriangledown$ \\
		 4 &&  0.382 &  4H  &  $22$     &  $\bigtriangleup\bigtriangleup\bigtriangledown\bigtriangledown$\\
		$\rightarrow$5 && 0.065 & 5R  & $14$    & $\bigtriangleup  \bigtriangledown \bigtriangledown \bigtriangledown\bigtriangledown$\\
		 6 &&  0.490 &  6H  &  $1122$ &  $\bigtriangleup  \bigtriangledown \bigtriangleup\bigtriangleup \bigtriangledown\bigtriangledown$\\
		$\rightarrow$7&& 0.08 & 7R  &  $1213$  &$\bigtriangleup  \bigtriangledown \bigtriangledown\bigtriangleup \bigtriangledown\bigtriangledown\bigtriangledown$\\
		8 && 0.267  & 8H  &$111122$ & $\bigtriangleup\bigtriangledown\bigtriangleup  \bigtriangledown \bigtriangleup\bigtriangleup \bigtriangledown\bigtriangledown$ \\
		8 && 0.130   & 8H  &  $112112$ & $\bigtriangleup\bigtriangledown\bigtriangleup  \bigtriangleup \bigtriangledown\bigtriangleup \bigtriangledown\bigtriangledown$\\
		$\rightarrow$9 && 0.062   & 9R  & $111213$ & $\bigtriangleup\bigtriangledown\bigtriangleup\bigtriangledown\bigtriangledown\bigtriangleup\bigtriangledown\bigtriangledown\bigtriangledown$ \\
		 9 && 0.056   & 9R  & $(12)_3$  &$\bigtriangleup\bigtriangledown\bigtriangledown\bigtriangleup\bigtriangledown\bigtriangledown\bigtriangleup\bigtriangledown\bigtriangledown$ \\
		 10 && 0.13  & 10H  &$11111122$  & $\bigtriangleup\bigtriangledown\bigtriangleup\bigtriangledown\bigtriangleup\bigtriangledown\bigtriangleup\bigtriangleup\bigtriangledown\bigtriangledown$ \\
		 10 &&  0.13   &  10H  & $11112112$  &  $\bigtriangleup\bigtriangledown\bigtriangleup\bigtriangledown\bigtriangleup\bigtriangleup\bigtriangledown\bigtriangleup\bigtriangledown\bigtriangledown$ \\
		 10 && 0.113  & 10H  & $112222$  & $\bigtriangleup\bigtriangledown\bigtriangleup\bigtriangleup\bigtriangledown\bigtriangledown\bigtriangleup\bigtriangleup\bigtriangledown\bigtriangledown$\\
		 10 && 0.056  & 10H  & $122122$  & $\bigtriangleup\bigtriangledown\bigtriangledown\bigtriangleup\bigtriangleup\bigtriangleup\bigtriangledown\bigtriangledown\bigtriangleup\bigtriangleup$ \\
		 11 &&  0.082  &  11R  & $11121212$ &  $\bigtriangleup\bigtriangledown\bigtriangleup\bigtriangledown\bigtriangledown\bigtriangleup\bigtriangledown\bigtriangledown\bigtriangleup\bigtriangledown\bigtriangledown$ \\
		 12 &&  0.062   & 12H  & $1111111122$  & $\bigtriangleup\bigtriangledown\bigtriangleup\bigtriangledown\bigtriangleup\bigtriangledown\bigtriangleup\bigtriangledown\bigtriangleup\bigtriangleup\bigtriangledown\bigtriangledown$ \\
		12 && 0.062 & 12H  & $1111112112$ & $\bigtriangleup\bigtriangledown\bigtriangleup\bigtriangledown\bigtriangleup\bigtriangledown\bigtriangleup\bigtriangleup\bigtriangledown\bigtriangleup\bigtriangledown\bigtriangledown$ \\
		 12 && 0.053  &  12H  & $ 11212212$  & $\bigtriangleup\bigtriangledown\bigtriangleup\bigtriangleup\bigtriangledown\bigtriangleup\bigtriangleup\bigtriangledown\bigtriangledown\bigtriangleup\bigtriangledown\bigtriangledown$ \\
		 12 && 0.054  & 12H  & $11122122$  &  $\bigtriangleup\bigtriangledown\bigtriangleup\bigtriangledown\bigtriangledown\bigtriangleup\bigtriangleup\bigtriangledown\bigtriangleup\bigtriangleup\bigtriangledown\bigtriangledown$ \\
		12 && 0.054  & 12H  & $11211222$  & $\bigtriangleup\bigtriangledown\bigtriangleup\bigtriangleup\bigtriangledown\bigtriangleup\bigtriangledown\bigtriangledown\bigtriangleup\bigtriangleup\bigtriangledown\bigtriangledown$ \\
		 12 &&  0.055  &  12H  &  $11112222$  &  $\bigtriangleup\bigtriangledown\bigtriangleup\bigtriangledown\bigtriangleup\bigtriangleup\bigtriangledown\bigtriangledown\bigtriangleup\bigtriangleup\bigtriangledown\bigtriangledown$ \\
		 12 &&  0.030  &  12H  &  $(11112)_2$  &  $\bigtriangleup\bigtriangledown\bigtriangleup\bigtriangledown\bigtriangleup\bigtriangleup\bigtriangledown\bigtriangleup\bigtriangledown\bigtriangleup\bigtriangledown\bigtriangledown$ \\
	\end{tabular}\label{tbl:hcpdhcp}
\end{table}

\bigskip

\newpage

\end{document}